\documentclass[a4paper,11pt]{article}
\usepackage{graphicx}
\usepackage[dvipdfm]{color}
\usepackage[varg]{txfonts}
\usepackage{indentfirst}
\begin{document}

\title{Constraint on R-parity violating MSSM at the one-loop level from CP-odd N-N interaction}

\author{Nodoka Yamanaka, Toru Sato, Takahiro Kubota\\
\\
{\it Department of Physics, Osaka University, Toyonaka, Osaka 560-0043, Japan}}
\date{}
\maketitle

\begin{abstract}
Minimal supersymmetric standard model with R-parity violation (RPVMSSM) contributes to the P-, CP-odd four-quark interaction. The P-, CP-odd four-quark interaction is constrained by the new $^{199}$Hg EDM experimental data. It is then possible to constrain R-parity violating (RPV) couplings from the $^{199}$Hg EDM data. In this talk, we analyze the RPV contribution to the P-, CP-odd four-quark interaction at the one-loop level to give constraints on RPV parameters.
\end{abstract}

\section{Introduction}
The standard model (SM) of particle physics is in good agreement with the current experimental data. There are however some phenomena which are difficult to explain in this framework, such as the matter abundance of our Universe. We need therefore a new physics beyond the SM to explain them.

As an efficient way to probe the new physics, we have the electric dipole moment (EDM) of $^{199}$Hg atom.
The EDM is an observable sensitive to the violation of parity and time-reversal symmetry.
The current experimental data of the $^{199}$Hg atomic EDM are very accurate ($d_{\rm Hg} < 3.1 \times 10^{-29} \mbox{e cm}$) \cite{griffith}.
The SM contribution to the $^{199}$Hg EDM is known to be very small ($\sim 10^{-34}$e cm), which renders it an excellent probe of new physics.

Among the new physics candidates, the supersymmetric extension of the SM, the minimal supersymmetric standard model (MSSM) is the most leading.
The supersymmetric extension of the SM allows baryon / lepton number violating interactions, so we must impose the {\it R-parity} ($R=(-1)^{3B - L  -2s}$) to forbid them.
This assumption is however totally {\it ad hoc}, and we must investigate the violation of R-parity phenomenologically \cite{chemtob}.

Faessler {\it et al.} gave constraints to the imaginary parts of some of the RPV couplings from the tree level analysis of the RPV contribution to the P-, CP-odd N-N interactions constrained by $^{199}$Hg EDM data \cite{faessler}.
On the experimental side, there was recently an update of the $^{199}$Hg EDM measurement by the group of Seattle \cite{griffith}. 
The purpose of this discussion is then to analyze the {\it one-loop level} contribution of the RPVMSSM to the P-, CP-odd four-quark interaction to obtain bounds on the RPV couplings.

Our talk is organized as follows. We first briefly review the R-parity violation and calculate its contribution to the atomic EDM at the one-loop level, then compare it with the new experimental data to obtain bounds on RPV couplings and finally summarize our results.

\section{RPV contribution to P-, CP-odd N-N interaction}
In the first step we construct the P-, CP-odd four-quark interaction at the one-loop level within RPVMSSM.
The RPV lagrangian used in this discussion is given as follows:
\begin{equation}
{\cal L} = 
 \lambda '_{ijk} (
 \tilde d^{\dag}_{Rk}\bar e_{i}^{c}P_{L}u_{j} + \tilde e_{Li}\bar d_{k}P_{L}u_{j} + \tilde u_{Lj}\bar d_{k}P_{L}e_{i}
- \tilde d_{Rk}^{\dag}\bar \nu _{i}^{c}P_{L}d_{j} - \tilde d_{Lj}\bar d_{k}P_{L}\nu _{i} - \tilde \nu _{i}\bar d_{k}P_{L}d_{j}
)
+{\rm h. c.}
\label{eq:rpvinteraction}
\end{equation}
with $P_{L}=(1-\gamma ^{5})/2$.
Here the indices $i,j,k =1,2,3$ denote the generations and the sum over them is taken.
Many of these RPV interaction terms are constrained phenomenologically \cite{chemtob}.

\begin{figure}
\begin{center}
\includegraphics[width=10cm]{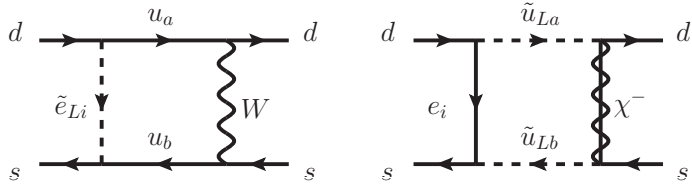}
\caption{Box diagrams contributing to P-, CP-odd four-quark interaction.}
\label{fig:box}
\end{center}
\end{figure}

\begin{figure}[h]
\begin{center}
\includegraphics[width=7cm]{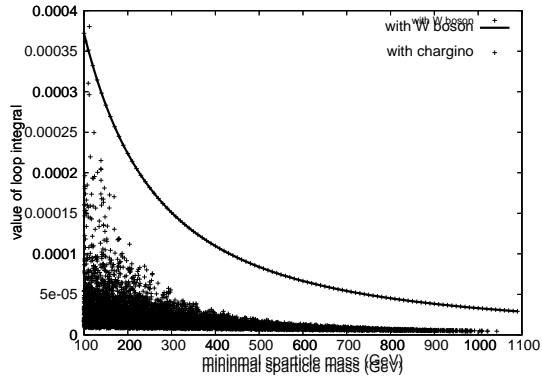}
\caption{Value of the loop integrals ($I$) of ${\cal M}_1$ (line) and ${\cal M}_2$ (dots). They are plotted as a function of the minimal sparticle mass in the loop.}
\label{fig:loop}
\end{center}
\end{figure}

By considering these RPV interactions, we can construct the one-loop contribution to the P-, CP-odd four-quark interaction. In enumerating the diagrams, we must exclude those which can be neglected or need not be taken into account:
Vertex corrections (renormalization of the tree level RPV coupling);
Diagrams with the same RPV couplings as the tree level (small, since constrained at the tree level \cite{faessler} );
Diagrams with Yukawa couplings of the 1st or 2nd generation (small contribution).
With these rules, there are only two contributing diagrams (plus their hermitian conjugates), which are shown in Fig. \ref{fig:box}.
The amplitude of the first diagram (with $W$ boson) and the second diagram (with chargino) are respectively
\begin{eqnarray}
{\cal M}_1 &=& i8 {\rm Im} (\lambda'^*_{ia1} \lambda'_{ib2}) V_{a1}V_{b2} \frac{G_F}{\sqrt{2}} I (m_W^2, m_{u_b}^2 , m_{\tilde e_{Li}}^2) \left[\bar d i\gamma_5 d \bar s s -\bar s i\gamma_5 s \bar d d + \cdots \right] ,\\
{\cal M}_2 &=& i8 {\rm Im} (\lambda'^*_{ia1} \lambda'_{ib2}) V_{a1}V_{b2} \frac{G_F}{\sqrt{2}} |Z_+^{1j}|^2 I (m_{\chi_j}^2, m_{\tilde u_{La}}^2 , m_{\tilde u_{Lb}}^2) \left[\bar d i\gamma_5 d \bar s s -\bar s i\gamma_5 s \bar d d \right] ,
\end{eqnarray}
where $I(x,y,z) = \frac{m_W^2}{4(4\pi )^2} \frac{1}{x-y} \left[ \frac{x}{z-x} \log \frac{z}{x} -\frac{y}{z-y} \log \frac{z}{y} \right] $, $V$ the CKM matrix elements, $a,b$ and $i$ are flavor indices. $\chi_j$ is the chargino with the sum over $j=1,2$ implied and $Z_+^{1j}$ is its mixing matrix.
We have neglected external and exchanged momenta.
The first and second diagrams have respectively one and three sparticles in the loop. The value of the loop integral of ${\cal M}_1$ and ${\cal M}_2$ is shown in Fig. \ref{fig:loop}. 
We do not know the sparticle masses, so we have calculated ${\cal M}_2$ by setting randomly 100GeV< $m_{\rm SUSY}$<1TeV with 10000 tries.
We see that ${\cal M}_2$ is generally smaller than ${\cal M}_1$.
These two contributions have the same RPV couplings with the {\it same sign}. From now, we focus only on the contribution of ${\cal M}_1$ in order to derive conservative upper bounds on RPV couplings.

The next step is the QCD calculation. We have used the factorization+PCAC method \cite{faessler} to evaluate the matrix element $
\langle p \pi^0 | \kappa_1 \frac{G_F}{\sqrt{2}} \bar d i \gamma_5 d \bar s s | p \rangle$ to give the P-, CP-odd four-quark contribution to the CP-odd $\pi N N $ vertex, where $\kappa_1 ( = 8 {\rm Im} (\lambda'^*_{ia1} \lambda'_{ib2}) V_{a1}V_{b2}  I (m_W^2, m_{u_b}^2 , m_{\tilde e_{Li}}^2)\   )$ is the four-quark coupling to be constrained. The CP-odd $\pi NN$ coupling is then
\begin{equation}
\bar g_{\pi NN} \approx 6 \times 10^{-6} \kappa_1
\end{equation}
Combined with the P-, CP-even $\pi NN$ vertex, we obtain the P-, CP-odd N-N interaction.
This P-, CP-odd N-N interaction is constrained by the $^{199}$Hg EDM data through the calculations of nuclear Schiff moment \cite{jesus} and atomic electron \cite{latha} of $^{199}$Hg atom. This gives $\bar g_{\pi NN} < 6.4 \times 10^{-13}$. 
We then obtain the following new limit to the product of RPV coupling, assuming sparticle masses $\approx 100 {\rm GeV}$
\begin{equation}
|{\rm Im} (\lambda'^*_{211} \lambda'_{232})|< 9 \times 10^{-4} .
\end{equation}

\section{Conclusion}
We have analyzed the RPV contribution to the P-, CP-odd N-N interaction and have given the new limit $|{\rm Im} (\lambda'^*_{211} \lambda'_{232})|< 9 \times 10^{-4}$ up to uncertainty of hadron matrix element .
Their lattice QCD study may therefore be interesting.

\section*{Acknowledgments}
This work is supported by JSPS, Grant-in-Aid for Scientific Research (C) 20540270.


\begin{thebibliography}{9}

\bibitem{griffith}
W. C. Griffith {\it et al.}, Phys. Rev. Lett. {\bf 102}, 101601 (2009).

\bibitem{chemtob}
M. Chemtob, Prog. Part. Nucl. Phys. {\bf 54}, 71 (2005).

\bibitem{faessler}
A. Faessler, T. Gutsche, S. Kovalenko and V. E. Lyubovitskij, Phys. Rev. D{\bf 73}, 114023 (2006).

\bibitem{jesus}
J. H.de Jesus and J. Engel, Phys. Rev. C{\bf 72},045503 (2005).

\bibitem{latha}
K. V. P. Latha, D. Angom, B. P. Das and D. Mukherjee, Phys. Rev. Lett. {\bf 103}, 083001 (2009).

\end{thebibliography}
\end{document}